\documentclass[12pt, preprint]{aastex}

\newcommand{\omc}{\hbox{$\omega$ Cen~}}
\newcommand{\omcp}{\hbox{$\omega$ Cen}}

\renewcommand{\deg}{\mbox{$^{\circ}$}}

\shorttitle{On the distance and reddening of the starburst galaxy IC10} 
\shortauthors{Sanna et al.}

\begin{document}
\title{On the distance and reddening of the starburst galaxy IC10\altaffilmark{1}}

\author{
N.\ Sanna\altaffilmark{2,3},
G.\ Bono\altaffilmark{2,3},
P.\ B.\ Stetson\altaffilmark{4},
M.\ Monelli\altaffilmark{5},
A.\ Pietrinferni\altaffilmark{6},
I.\ Drozdovsky\altaffilmark{5},
F.\ Caputo\altaffilmark{3},
S.\ Cassisi\altaffilmark{6}, 
M.\ Gennaro\altaffilmark{7},
P.\ G.\ Prada Moroni \altaffilmark{7,8},
R.\ Buonanno\altaffilmark{2},
C.\ E.\ Corsi\altaffilmark{3},
S.\ Degl'Innocenti\altaffilmark{7,8}
I.\ Ferraro\altaffilmark{3},
G.\ Iannicola\altaffilmark{3},
M.\ Nonino\altaffilmark{9}, 
L.\ Pulone \altaffilmark{3},  
M.\ Romaniello\altaffilmark{10}, and  
A.\ R. \ Walker\altaffilmark{11}
}

\altaffiltext{1}
   {Based on observations collected with the ACS and with the WFPC2 
on board of the HST.}
  \altaffiltext{2}{UniToV, via della Ricerca Scientifica 1, 00133 Rome, Italy; 
Nicoletta.Sanna@roma2.infn.it}
  \altaffiltext{3}{INAF--OAR, via Frascati 33, Monte Porzio Catone, Rome, Italy}
  \altaffiltext{4}{DAO--HIA, NRC, 5071 West Saanich Road,
      Victoria, BC V9E 2E7, Canada} 
  \altaffiltext{5}{IAC, Calle Via Lactea, E38200 La Laguna, Tenerife, Spain}
  \altaffiltext{6}{INAF--OACTe, via M. Maggini, 64100 Teramo, Italy}
  \altaffiltext{7}{Univ. Pisa, Largo B. Pontecorvo 2, 56127 Pisa, Italy}
  \altaffiltext{8}{INFN, Sez. Pisa, via E. Fermi 2, 56127 Pisa, Italy}
  \altaffiltext{9}{INAF--OAT, via G.B. Tiepolo 11, 40131 Trieste, Italy}
  \altaffiltext{10}{ESO, Karl-Schwarzschild-Str. 2, 85748 Garching bei Munchen, Germany}
 \altaffiltext{11}{CTIAO--NOAO, Casilla 603, La Serena, Chile}

%\date{\centering drafted \today\ / Received / Accepted }

\begin{abstract}
We present deep and accurate optical photometry of the Local Group 
starburst galaxy IC10. The photometry is based on two sets of images 
collected with the Advanced Camera for Surveys and with the Wide Field 
Planetary Camera 2 on board the Hubble Space Telescope. We provide new 
estimates of the Red Giant Branch tip (TRGB) magnitude, 
$m_{F814W}^{TRGB}$=$21.90$$\pm$$0.03$, and of the reddening, 
E(B-V)=$0.78$$\pm$$ 0.06$, using field stars in the Small Magellanic Cloud 
(SMC) as a reference. Adopting the SMC and two globulars, \omc and 47~Tuc, as
references we estimate the distance modulus to IC10: independent calibrations
give weighted average distances of $\mu$$=$$24.51$$\pm$$0.08$ (TRGB) and  
$\mu$$=$$24.56$$\pm$$0.08$ (RR Lyrae). We also provide a new theoretical
calibration for the TRGB luminosity, and using these predictions we find 
a very similar distance to IC10 ($\mu$$\approx$$24.60$ $\pm$$0.15$). 
% E
These results suggest that IC10 is a likely member of the M31 subgroup. 
\end{abstract}


%\maketitle

%%%%%%%%%%%%%%%%%%%%%%%%%%%%%%%%%%%%%%%%%%%%%%%%%%%%%%%%%%%%%%%%%%%%%
\section{Introduction}

The dwarf galaxy IC10 is a very interesting stellar system in the Local Group. 
It has been classified as an Ir IV by van~den~Bergh (1999), but it has
also been suggested that 
% F
it is the only analog of a post-starburst dwarf galaxy in the Local Group 
(Gil de Paz et al.\ 2003).  
This means that IC10 is a very good laboratory to investigate
episodic star formation over a broad time interval (Hunter 2001; Demers et al.\
2004).  
% B
Moreover, spectroscopic estimates based on HII regions indicate that
the present-day interstellar medium metal abundance is $[Fe/H]\sim-0.71\pm0.10$,
(Garnett et al.\ 1990). We cannot exclude the possibility of a range
of metallicity in IC10, but comparison with stellar
isochrones (Hunter 2001) also indicates a mean metallicity of
$[Fe/H]\sim-0.7$. This metal content is very similar to that of the
Small Magellanic Cloud (SMC; $[Fe/H]\approx -0.7$, Zaritsky et al.\ 1994,
based on HII regions; $[Fe/H]\sim-0.75\pm0.08$, Romaniello et al.\ 2008,
based on classical Cepheids).
% H 
% Therefore, its stellar content can help to reveal the role that 
% the environment has on the evolutionary properties and the chemical 
% evolution of metal-intermediate systems.  
  
However, IC10 is located at very low Galactic latitude (l=119\deg0, b=-3\deg3),
and is therefore affected by significant foreground extinction. This
circumstance has made estimates of both reddening and distance quite
difficult and controversial. 
% Dating back to its discovery (Mayall 1935), 
Distance determinations based on different standard candles cover
% I 
a wide range, from closer than M31---$d=0.5$ Mpc based on the Red 
Giant Branch tip (TRGB, Sakai et al.\ 1999)---to well outside the Local Group, 
$d=1.8$ Mpc based on the luminosity function (LF) of planetary nebulae 
(Jacoby \& Lesser 1981). 
A similar disparity 
is found among the reddening determinations, with estimates ranging from
$E(B-V)\sim 0.8$ based on spectra of HII regions (Richer et al.\ 2001), to
$E(B-V)\sim 1.2$ based on classical Cepheids (Sakai et al.\ 1999). Moreover, a
spread in distance and reddening values can be found even when using the same 
distance indicator and stellar tracer (see Table~1, in Demers et al.\ 2004).
Therefore, we still lack firm constraints on the random and systematic errors
affecting these important parameters. 

% J 
% To overcome some of these problems, we intend to provide reddening estimates 
% based on empirical calibrators and both relative and absolute distances. 

%-------------------------------------------------------------------------
\section{Observations and data reduction}

Our photometric catalog is based on archival Hubble Space Telescope (HST) 
data sets
% K 
collected with both the Advanced Camera for Surveys (ACS, pointings  $\alpha$
and $\beta$) and with the Wide Field Planetary Camera 2 (WFPC2, pointing
$\gamma$)\footnote{The pointing coordinates (J2000) are: RA=00~20~19, 
DEC=59~18~24 ($\alpha$, GO--9683, PI: F.E. Bauer); RA=00~20~14, DEC=59~20~23 
($\beta$, GO--10242, PI: A.A. Cole); RA=00~20~28, DEC=59~18~00 ($\gamma$, 
GO--6406, PI: D.A.  Hunter).}.
Pointing $\alpha$ is located at the galaxy center and consists of six $F606W$ and
six $F814W$-band images of 360 sec each. Pointing $\beta$ is located $\sim 2'$
NW from the galaxy center and includes 32 $F555W$-band images of 620 sec each
and 16 $F814W$-band images of 595 sec each.  Pointing $\gamma$ is again located
at the galaxy center and includes ten $F555W$ and ten $F814W$-band images of
1400 sec each. 
% A 
Pointings $\alpha$ and $\beta$ (ACS) overlap by about one chip,
while pointing $\gamma$ (WFPC2) almost entirely overlaps with pointing
$\alpha$.
We combined the ACS images using an updated version of the {\tt MultiDrizzle} 
package (Koekemoer et al.\ 2002), which provides an automated method for 
correcting  distortion and combining dithered images. 
%{\tt Multidrizzle} also corrects for gain and bias offsets between 
%the WFC chips, identifies and removes cosmic rays and cosmetic defects. 
The WFPC2 images were 
prereduced using the HST pipeline. Initial photometry on individual 
% K bis  
images was performed with {\tt DAOPHOT IV}, followed by simultaneous 
photometry over the 80 images with {\tt ALLFRAME} (Stetson, 1994). We ended
up with a catalog including $\sim 720,000$ stars with at least one measurement
in each of two different bands. The ACS data in the $F555W$ and $F814W$ bands
were transformed into the VEGAMAG system following Sirianni et al.\ (2005).  To
provide a homogeneous photometric catalog the $F606W$-band images collected with
the ACS were transformed into the $F555W$-band using local standards. The same
approach was adopted to transform the $F555W$ and the $F814W$ images collected
with WFPC2 to the corresponding ACS systems. On average the precision of the
above transformations is $F555W$-$F606W$(ACS)= -0.006 $\pm$ 0.049, 
$F555W$(ACS)-$F555W$(WFPC2)= 0.009 $\pm$ 0.054, 
$F814W$(ACS)-$F814W$(WFPC2)= 0.003 $\pm$ 0.050 mag (Sanna et al.\ 2008, 
in preparation). To overcome possible changes in the internal reddening when
moving from the center toward the external regions of the galaxy we split the
final catalog into two different regions.  Region C is located at the galaxy
center, while the field lying at a radial distance larger than two arcminutes is
our region E.  Data plotted in Fig.~1 show that the current photometry ranges from
very bright main sequence stars ($F814W\sim 21$, $F555W$-$F814W$$\sim 1$) to 
faint RG stars ($F814W\sim 25.5$, $F555W$-$F814W$$\sim 2$) with very
good precision.    

%%%%%%%%%%%%%%%%%%%%%%%%%%%%%%%%%%%%%%%%%%%%%%%%%%%%%%%%%%%%%%%%%%%%%%%%%%%%%
\section{Results and discussion}

To estimate the TRGB in IC10 we adopted the approach used in Bono et al.\
(2008, hereinafter B08). 
% L
In particular, we selected stars with $21.5\le$ $F814W$ $\le 22.4$ and
$F555W-F814W$$\ge 2.8$. 
Note that the stellar samples in the central and external 
regions include, respectively, $\sim 3,000$ and  $\sim 2,400$ stars within one
$F814W$-band magnitude of the tip. This is more than an order of magnitude larger
than the number required for a robust detection of the TRGB (Madore
\& Freedman 1995).  The top panels of Fig.~2 show a well defined jump 
in the star counts for $m_{F814W}\sim 21.90$, which we take to mark the position
of the TRGB.  This identification is supported by the smoothed LF  
obtained with a Gaussian kernel having a standard deviation equal to
the photometric error (Sakai et al.\ 1996; middle panels).  Finally,
the bottom panels show the response of the edge detector, a four-point 
Sobel filter convolved with the smoothed LF. The dashed vertical lines 
mark our detection of the TRGB at $m_{F814W}^{TRGB}$= $21.90\pm0.025$ 
for the central field (left) and $m_{F814W}^{TRGB}$= $21.89\pm0.025$ 
for the external field (right). 
% L bis  
The error in these TRGB estimates is given by the bin size adopted in
the LF (Fig.~2, top panels).

We adopted empirical calibrators to determine the reddening and the
distance to IC10. This approach is minimally affected by uncertainties in the 
transformation of theoretical predictions into the observational plane. 
Moreover, it provides robust estimates of {\it relative\/} distances and 
reddenings. The main drawback is that the empirical calibrators require 
%similar mean metal contents (or at least trustworthy metallicity corrections).
similar mean metallicities (or at least trustworthy metallicity corrections).
To estimate the reddening we used the accurate  
photometry of the region located around the SMC cluster NGC~346 recently 
provided by Sabbi et al.\ (2007). This data set offers several advantages: 
{\em i)}~the data were collected with ACS in the same  $F555W$, $F814W$-bands; 
{\em ii)}~the stellar populations in this region span a very broad age range, 
namely from the few Myrs of NGC~346 to the several Gyrs ($t=4.5$ Gyr) of the 
intermediate-age cluster BS90; {\em iii)}~the region is characterized by 
modest reddening (E(B-V)=0.08) and is minimally contaminated by Galactic field 
stars; 
% B
{\em iv}) the mean metallicity is very similar to the mean metallicity of IC10.  

The relative reddening was estimated by moving the young SMC MS until it 
matched the bluest MS stars in IC10 to account for possible differential 
reddening. Fortunately enough, the distribution 
of these stars in the $F814W$, $F555W$-$F814W$ CMD is almost vertical 
and with the RGs they form a ``V'' shape.  Therefore, this approach 
minimally depends on the adopted distance. We found a relative reddening of 
$\delta$$E($F555W$-$F814W$)$=$1.04\pm 0.05$ for both the central and the 
external region of IC10. 
% M 
This indicates that the reddening in the two fields is, within the 
errors, the same. 
Using the analytical relation for stellar extinction provided by Cardelli et
al.\ (1989) we found that the selective absorptions in the two ACS filters are: 
$A_{F555W}=1.03\times A_V$ and $A_{F814W}=0.55\times A_V$.  If we assume an  
SMC reddening of E(B-V)=0.08 (Sabbi et al.\ 2007), this means a mean
reddening of E($F555W$-$F814W$)=$1.16\pm 0.06$ for IC10, and in turn
$E(B-V)$=0.78$\pm$0.06.  The error budget includes generous estimates on 
the relative reddening ($\pm 0.05$), on the SMC reddening ($\sim 0.02$) 
and on the reddening law (Fitzpatrick 1999). 
The left panel of Fig.~3 shows the comparison between the SMC (green dots) and
IC10 (gray dots) using the above differential reddening and relative distance
based on the TRGB (see {\em infra}).  The new reddening estimate is in very good
agreement with the reddening estimates based on spectra of HII regions
(E(B-V)=0.77 $\pm$ 0.07, Richer et al.\ 2001), TRGB stars 
(E(B-V) $\sim$ 0.85, Sakai et al.\ 1999), carbon stars (E(B-V) $\sim$ 0.79, 
and a change across the field of $\sim$ 0.38 mag, Demers et al.\ 2004), 
and Wolf-Rayet stars
(E(B-V)=0.75-0.80, Massey \& Armandroff 1995).  We do not support a 
reddening as high as was estimated from classical Cepheid variables.

We also estimated the relative distance 
between IC10 and the SMC. For the SMC, we adopted the TRGB estimate  
provided by Cioni et al.\ (2000), i.e., $m_{I}(SMC)=14.95\pm0.03$. 
With the reddening quoted above for SMC (E(B-V)=0.08), the color-metallicity 
relation for TRGB stars provided by Bellazzini et al.\ (2001), and the 
transformations from the Cousins $I$-band to the ACS VEGAMAG system 
by Sirianni et al.\ (2005), we found $m_{F814W,0}(SMC)=14.81\pm0.05$. 
Using the above reddening, we found that $m_{F814W,0}(IC10)=20.57\pm0.07$ 
and the error budget includes the same sources of uncertainty as before, 
plus an additional uncertainty of $\pm 0.02$~mag in the photometric
calibration.  Therefore, the relative distance is $\Delta \mu= 5.76\pm
0.09$.  The arrow plotted in the middle panel of Fig.~3 shows that the 
magnitudes of TRGB stars in SMC and in IC10 agree within the uncertainties.
  
To further constrain the relative distance to IC10, we also compared it to
the Galactic Globular Cluster (GGC) 47~Tuc. The reason is threefold: 
{\em i)}~accurate spectroscopic measurements indicate that its iron 
abundance is very similar to IC10, i.e., [Fe/H]=$-0.66\pm0.04$ 
(Gratton et al.\ 2003); {\em ii)} it is minimally affected 
by reddening: (E(B-V)=$0.04\pm0.02$, Salaris et al.\ 2007); 
{\em iii)} an accurate estimate of the TRGB has been recently 
provided by B08. By transforming the Cousins $I$-band 
into the ACS VEGAMAG system (Sirianni et al.\ 2005), we found that the 
TRGB in 47Tuc is located at $m_{F814W,0}(47Tuc)=9.40\pm0.08$.  The error 
budget accounts for uncertainties in the reddening, in the TRGB detection 
and in the photometric transformation. Therefore, the relative distance 
is $\Delta \mu= 11.17\pm 0.11$. A glance at the data plotted in the 
middle panel of Fig.~3 shows very good agreement between the magnitudes 
of TRGB stars in 47Tuc and IC10. Moreover, the RGs in 47Tuc also attain 
very similar colors to RGs in IC10, thus supporting the reddening 
(and metallicity) estimate.    

We also estimated the relative distance between IC10 and the GGC \omcp. 
Again, the reason is threefold:
{\em i)} it is the most massive GGC and it hosts a sizable sample
of bright RG stars ($\sim 220$) close to TRGB, so the TRGB 
detection is particularly robust;  {\em ii)} accurate distance estimates are
available using several different standard candles (Del Principe et al.\ 2006); 
{\em iii)} an accurate estimate of the TRGB has been recently provided 
by B08. We found that the TRGB in \omc is located at 
$m_{F814W,0}(\omega Cen)=9.73\pm0.09$. The error budget accounts 
for uncertainties in the reddening ($E(B-V)=0.11\pm0.02$), 
in the TRGB detection, and in the photometric transformation. 
Moreover, we also accounted for the difference in metallicity 
between \omc ([Fe/H]$\sim -1.7$, metal-poor peak, Johnson et al.\ 2008) 
and IC10, and their uncertainties (B08). The relative distance is then
$\Delta \mu= 10.84\pm 0.11$. The right panel of Fig.~3 shows that 
the magnitudes of TRGB stars in \omc and in IC10 agree quite well. The 
% O 
bright 
RGs in \omc are, as expected, bluer when compared with RGs in IC10. 

To evaluate the true distance to IC10, we used the TRGB 
calibrations provided by Bellazzini et al.\ (2004) and 
Lee et al.\ (1993).  For the SMC, we found a true modulus of 
$\mu$= $18.77\pm 0.13$ or $\mu$= $18.73\pm 0.12$; $\mu$=$18.75\pm0.09$ 
is the weighted average. Thus, the true modulus of IC10, based on the 
TRGB scale of SMC, is $\mu$= $24.51\pm 0.13$. Using the same 
TRGB calibrations we found for 47~Tuc a true modulus of 
$\mu$= $13.36\pm 0.14$ and $\mu$= $13.32\pm 0.12$ ($\mu$=$13.34\pm0.10$,
weighted average). Thus, the modulus of IC10, based on the 
TRGB scale of 47~Tuc, is $\mu$= $24.51\pm 0.15$. Finally, using the same TRGB
calibrations the true moduli for \omc are: $\mu$= $13.67\pm 0.14$ 
and $\mu$= $13.65\pm 0.13$ ($\mu$=$13.66\pm0.10$, weighted average). 
Therefore, the modulus of IC10, based on the TRGB scale of \omc, is $\mu$=
$24.50\pm 0.15$. 

On the other hand, if we use the $K$-band Period-Luminosity (PL)  
relations provided by Del Principe et al.\ (2006) and 
Sollima et al.\ (2008), the true modulus of 47~Tuc based on the variable 
V9 is $\mu= 13.38\pm 0.06$ and $\mu= 13.47\pm 0.11$ ($\mu=13.40\pm 0.05$, 
weighted average)\footnote{We did not estimate the distance to IC10 
using the SMC RR Lyrae, since NIR magnitudes for these objects
are not available.}.  
Therefore, the modulus of IC10, based on the RR Lyrae scale, is 
$\mu= 24.57\pm 0.12$. Using the same $K$-band PL relations,
the true modulus of \omc is $\mu= 13.70\pm 0.06$ or 
$\mu= 13.75\pm 0.11$ ($\mu=13.71\pm 0.05$, weighted average).  
Therefore, the modulus of IC10, based on the RR Lyrae scale 
to \omc, is $\mu= 24.55\pm 0.12$.  
The weighted average true moduli of IC10 agree within 1$\sigma$ and are: $\mu=
24.51\pm 0.08$ (TRGB scale) and $\mu= 24.56\pm 0.08$ (RR Lyrae scale). 
These estimates agree within 1$\sigma$ with previous distance determinations
based on carbon stars ($\mu= 24.35\pm 0.11$, E(B-V)=0.79, Demers et al.\ 2004);
on optical ($\mu= 24.59\pm 0.30$, E(B-V)=0.97, Saha et al.\ 1996) and NIR ($\mu=
24.57\pm 0.21$, E(B-V)=0.8, Wilson et al.\ 1996) Cepheid PL 
relations; and on the TRGB ($\mu= 24.48\pm 0.08$, E(B-V)=0.95, Vacca et al.\
2007). However, we found differences of $\approx$ 0.4--0.5 mag with 
respect to the moduli estimated from Wolf-Rayet stars ($\mu= 24.9$,
E(B-V)=0.75--0.80, Massey \& Armandroff 1995) and red supergiants ($\mu=
23.86\pm 0.12$, E(B-V)=$1.05\pm 0.10$, Borissova et al.\ 2000).   
% P 
This indicates that the quoted distance estimates might be affected by
systematic errors. 
   
As a final test of the systematic errors that might affect the distance
estimates to IC10, we calculated new theoretical TRGB calibrations. We adopted
the homogeneous set of cluster isochrones for scaled-solar abundances provided
by Pietrinferni et al.\ (2004); in particular, we adopted isochrones 
with t=12 Gyrs, mass-loss $\eta$=0.4, and iron abundances from $-2.3 \lesssim$ 
[M/H] $\lesssim -0.5$ (see the URL {\tt http://www.oa-teramo.inaf.it/basti}). 
However, theory was transformed into the observational plane 
using only scaled-solar atmospheres based on {\tt ATLAS9} models
(Castelli et al.\ 2003). We linearly extrapolated the surface gravity by 
$\sim$0.1 dex and the surface temperature by $\sim$100 K beyond 
the available range for the most metal-rich structures. 
To account for the new electron-conduction opacities the predicted 
TRGB magnitudes were dimmed by 0.075 mag (Cassisi et al.\ 2007). 
Moreover, to account for uncertainties in the color-temperature 
transformations (Bellazzini 2008) we also adopted scaled-solar 
atmospheres based on {\tt PHOENIX} (Brott et al.\ 2005).

The new ATLAS9 and PHOENIX TRGB calibrations agree well (Fig.~4, bottom panel):
over most of the metallicity range differences are much less than 0.1 mag, but
they do become perceptibly larger for the most metal-rich structures.
% C 
The new TRGB calibrations also agree (Fig.~4, top panel), within 
1$\sigma$ (see the error bars), with empirical TRGB calibrations 
(Lee et al. 1993; Bellazzini et al.\ 2004; Rizzi et al.\ 2007).
Predicted $M_I^{TRGB}$ magnitudes are still systematically brighter than
observed ones, but the difference is a factor of two smaller than previous
TRGB calibrations. The explanation of this mild discrepancy requires new
theoretical and empirical investigations. 
% D 
We performed spline fits of the theoretical TRGB luminosities\footnote{The 
predicted TRGB luminosities were estimated for the following metallicities: 
[M/H]= -2.267, -1.790,  -1.488,  -1.266, -0.963, -0.659, -0.481. The set of 
isochrones for [M/H]= -0.481 was specifically computed for this project. 
The anchor points we adopted for the spline  fit are: 
$M_I^{TRGB}(ATLAS9)$= -4.085, -4.137, -4.143, -4.129, -4.088, -4.034, -3.977; 
$M_I^{TRGB}(PHOENIX)$= -4.066, -4.110, -4.108, -4.106, -4.069, -4.012, -3.870;
$M_{F814W}^{TRGB}(ATLAS9)$= -4.089, -4.140, -4.143, -4.130, -4.085, -4.018, -3.947;
$M_{F814W}^{TRGB}(PHOENIX)$= -4.104, -4.149, -4.153, -4.138, -4.106, -4.034, -3.894.}
in both
$I$ and $F814W$-band, and from these---by assuming a metal content for IC10
of [M/H]=-0.70---we derived true distance moduli of $\mu = 24.59\pm 0.13$ and
$\mu = 24.60\pm0.15$, respectively. 
The error
budget includes uncertainties in the theoretical predictions ($\pm0.10$~mag),
the TRGB measurement, and in the metal content ($\pm0.20$~dex). These estimates
are in very good agreement with the above distances based on empirical
calibrators, thus further supporting the accuracy of the current theoretical
calibrations.  

The above distances indicate that IC10 is located, within the errors, at the
same distance as M31, which has true moduli based on the TRGB ranging from
$24.47\pm 0.11$ to $24.37\pm 0.08$ (Rizzi et al.\ 2007), while those based on
Cepheids range from $24.38\pm 0.05$ (Sakai et al.\ 2004) to $24.32\pm 0.12$
(Vilardell et al. 2007; Tammann et al.\ 2008). The new distances are also 
in remarkable agreement with the ``rotational parallax'' for M33 
based on water masers ($ 805$$\pm$$ 37$ vs $730$$\pm$$168$ kpc) 
by Brunthaler et al.\ (2005).   
% E 
This indicates that IC10 is at the same distance as M31 and M33, 
and in turn that it is a likely member of the M31 subgroup.

\acknowledgments
It is a real pleasure to thank E. Sabbi for sending us her SMC data in 
electronic form. This paper was partially supported by PRIN-INAF (PI: 
M. Bellazzini). 

%--------------------------------------------------------------------------------------

\clearpage

%%%%%%%%%%%%%%%%%%%%%%%%%%%%%%%%%%%%%%%%%%%%%%%%%%%%%%%%%%%%%%%%%%%%%%%%%%%%%%%%%%%%%
% 			fig 1 
%%%%%%%%%%%%%%%%%%%%%%%%%%%%%%%%%%%%%%%%%%%%%%%%%%%%%%%%%%%%%%%%%%%%%%%%%%%%%%%%%%%%%
\begin{figure}[!ht]
\begin{center}
\label{fig1}
\includegraphics[height=0.25\textheight,width=0.45\textwidth]{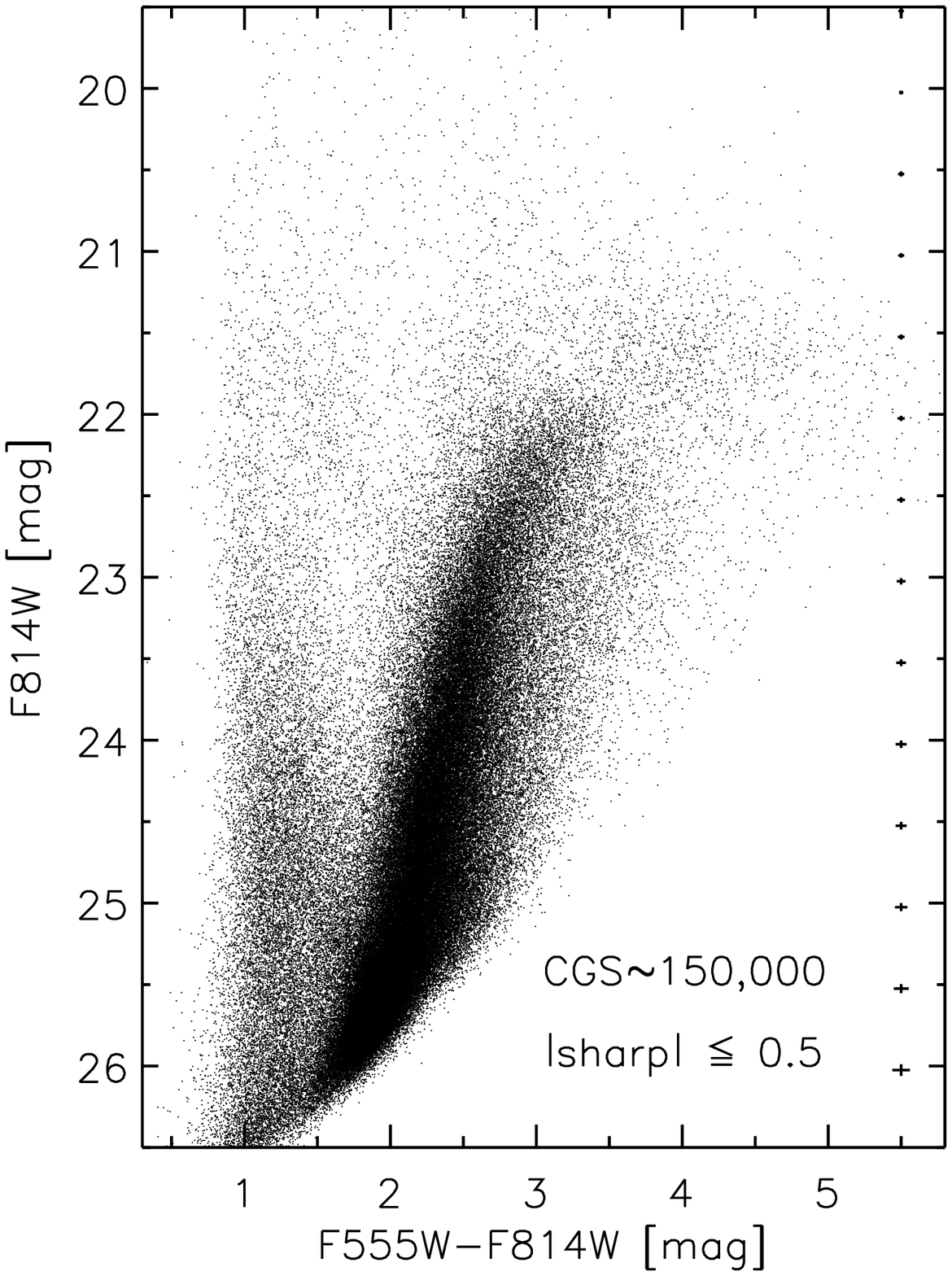}
\vspace*{-0.05truecm}
\caption{
% A 
$F814W$, $F555W$-$F814W$ CMD of IC10 based on data collected with HST
(fields C and E). Using different selection criteria ({\em sharpness},
{\em separation}, photometric error) we ended up with $\approx$ 150,000
candidate galaxy stars (CGS). The error bars on the right display the
mean intrinsic error.
}
\end{center}
\end{figure}

%%%%%%%%%%%%%%%%%%%%%%%%%%%%%%%%%%%%%%%%%%%%%%%%%%%%%%%%%%%%%%%%%%%%%%%%%%%%%%%%%%%%%
% 			fig 2 
%%%%%%%%%%%%%%%%%%%%%%%%%%%%%%%%%%%%%%%%%%%%%%%%%%%%%%%%%%%%%%%%%%%%%%%%%%%%%%%%%%%%%
\begin{figure}[!ht]
\begin{center}
\label{fig1}
\includegraphics[height=0.25\textheight,width=0.45\textwidth]{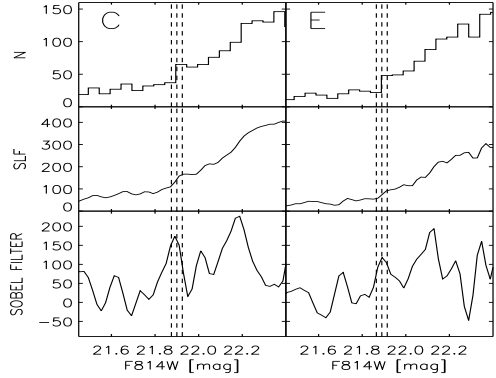}
\vspace*{-0.05truecm}
\caption{Left -- Top -- $F814W$-band LF for the bright end of
the RGB in the central region. Middle -- Smoothed LF of the
same RGB region obtained using a Gaussian kernel with standard deviation equal
to the photometric error. Bottom -- Response of the four-point Sobel filter to 
the smoothed LF. The dashed vertical lines indicate the position of the TRGB
% R 
and its 1$\sigma$ error. 
Right -- Same as the left, but for bright RGB stars in the external region.  
}
\end{center}
\end{figure}

%%%%%%%%%%%%%%%%%%%%%%%%%%%%%%%%%%%%%%%%%%%%%%%%%%%%%%%%%%%%%%%%%%%%%%%%%%%%%%%%%%%%%
% 			fig 3 
%%%%%%%%%%%%%%%%%%%%%%%%%%%%%%%%%%%%%%%%%%%%%%%%%%%%%%%%%%%%%%%%%%%%%%%%%%%%%%%%%%%%%
\begin{figure}[!ht]
\begin{center}
\label{fig2}
\includegraphics[height=0.4\textheight,width=0.5\textwidth]{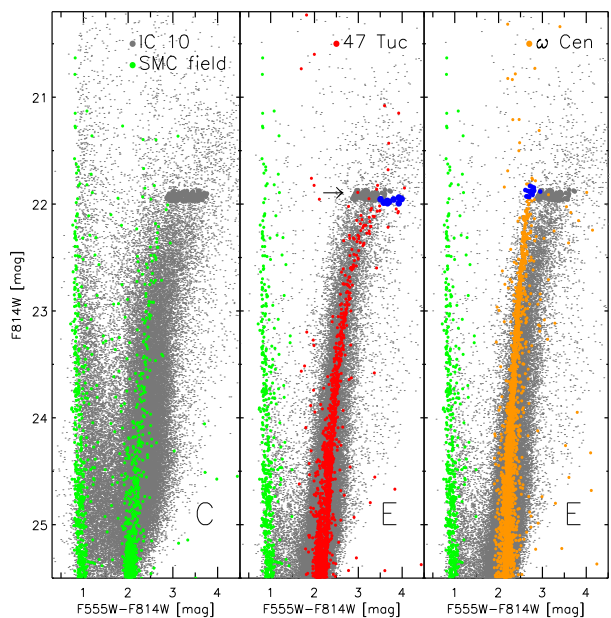}
\vspace*{-0.15truecm}
\caption{Left -- Comparison in the $F814W$, $F555W$-$F814W$ CMD between 
IC10 stars located in the central region (C, gray dots) and MS, RG 
stars in the SMC cluster NGC~346 (green dots). 
The large gray circles mark TRGB stars in IC10. Middle -- Same as the 
left, but the comparison is between IC10 stars located in the external region 
(E, gray dots) and RG stars in 47Tuc (red dots). The large blue circles mark 
the TRGB stars in 47Tuc. 
The horizontal arrow marks the position of TRGB stars in SMC according to 
the measurement by Cioni et al.\ (2000). 
Right -- Same as the middle, but the comparison is between IC10 stars in the 
external region and RG stars in \omc (orange dots). The large blue dots show 
TRGB stars in \omc according to B08.   
}
\end{center}
\end{figure}

%%%%%%%%%%%%%%%%%%%%%%%%%%%%%%%%%%%%%%%%%%%%%%%%%%%%%%%%%%%%%%%%%%%%%%%%%%%%%%%%%%%%%
% 			fig 4 
%%%%%%%%%%%%%%%%%%%%%%%%%%%%%%%%%%%%%%%%%%%%%%%%%%%%%%%%%%%%%%%%%%%%%%%%%%%%%%%%%%%%%
\begin{figure}[!ht]
\begin{center}
\label{fig4}
\includegraphics[height=0.25\textheight,width=0.45\textwidth]{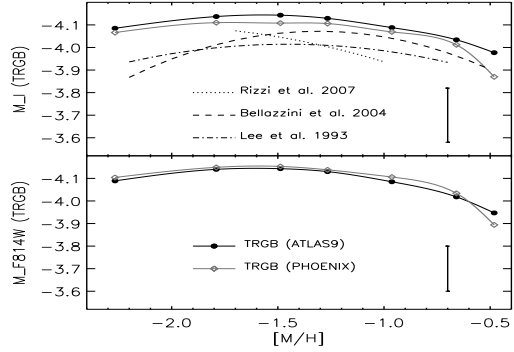}
\vspace*{-0.15truecm}
\caption{Top -- Comparison between theoretical and empirical 
$M_I^{TRGB}$ calibrations versus metallicity. Circles and 
diamonds display predictions according to {\tt ATLAS9} and {\tt PHOENIX} 
atmosphere models. 
% T 
The error bar shows the typical uncertainty on epirical calibrations 
(Bellazzini 2008). Bottom -- Same as the top, but the theoretical predictions 
for the TRGB were transformed into the $F814W$-band. The error bar shows 
current uncertainty on theory.    
}
\end{center}
\end{figure}

\end{document}